\pdfoutput=1
\documentclass[%
aip,
rsi,%
amsmath,amssymb,
reprint,%
]{revtex4-1}

\usepackage{graphicx}
\usepackage[space]{grffile}
\usepackage{dcolumn}
\usepackage{bm}
\usepackage{float}
\begin{document}

\preprint{AIP/123-QED}

\title[]{Quantitative comparison of the magnetic proximity effect in Pt detected by XRMR and XMCD}

\author{Dominik Graulich}
\email{dgraulich@physik.uni-bielefeld.de}
\affiliation{ 
	Center for Spinelectronic Materials and Devices, Department of Physics, Bielefeld University, Universit\"atsstra\ss{}e 25, 33615 Bielefeld, Germany\looseness=-1}
\author{Jan Krieft}
\affiliation{ 
	Center for Spinelectronic Materials and Devices, Department of Physics, Bielefeld University, Universit\"atsstra\ss{}e 25, 33615 Bielefeld, Germany\looseness=-1}%
\author{Anastasiia Moskaltsova}
\affiliation{ 
	Center for Spinelectronic Materials and Devices, Department of Physics, Bielefeld University, Universit\"atsstra\ss{}e 25, 33615 Bielefeld, Germany\looseness=-1}%
\author{Johannes Demir}
\affiliation{ 
	Center for Spinelectronic Materials and Devices, Department of Physics, Bielefeld University, Universit\"atsstra\ss{}e 25, 33615 Bielefeld, Germany\looseness=-1}%
\author{Tobias Peters}
\affiliation{ 
	Center for Spinelectronic Materials and Devices, Department of Physics, Bielefeld University, Universit\"atsstra\ss{}e 25, 33615 Bielefeld, Germany\looseness=-1}%
\author{Tobias Pohlmann}
\affiliation{ 
	Center of Physics and Chemistry of New Materials, Department of Physics, Osnabr\"uck University, 
	Barbarastra\ss{}e 7, 49076 Osnabr\"uck, Germany\looseness=-1}
\affiliation{ 
	Deutsches Elektronen-Synchrotron DESY, Notkestra\ss{}e 85, 22607 Hamburg, Germany\looseness=-1}
\author{Florian Bertram}
\affiliation{ 
	Deutsches Elektronen-Synchrotron DESY, Notkestra\ss{}e 85, 22607 Hamburg, Germany\looseness=-1}
\author{Joachim Wollschl\"ager}
\affiliation{ 
	Center of Physics and Chemistry of New Materials, Department of Physics, Osnabr\"uck University, 
	Barbarastra\ss{}e 7, 49076 Osnabr\"uck, Germany\looseness=-1}
\author{Jose R. L. Mardegan}
\affiliation{ 
	Deutsches Elektronen-Synchrotron DESY, Notkestra\ss{}e 85, 22607 Hamburg, Germany\looseness=-1}%
\author{Sonia Francoual}
\affiliation{ 
	Deutsches Elektronen-Synchrotron DESY, Notkestra\ss{}e 85, 22607 Hamburg, Germany\looseness=-1}%
\author{Timo Kuschel}%
\affiliation{ 
	Center for Spinelectronic Materials and Devices, Department of Physics, Bielefeld University, Universit\"atsstra\ss{}e 25, 33615 Bielefeld, Germany\looseness=-1}
\date{\today}

\begin{abstract}
X-ray resonant magnetic reflectivity (XRMR) allows for the simultaneous measurement of structural, optical and magnetooptic properties and depth profiles of a variety of thin film samples. However, a same-beamtime same-sample systematic quantitative comparison of the magnetic properties observed with XRMR and x-ray magnetic circular dichroism (XMCD) is still pending. Here, the XRMR results (Pt L$_{3}$ absorption edge) for the magnetic proximity effect in Pt deposited on the two different ferromagnetic materials Fe and Co$_{33}$Fe$_{67}$ are compared with quantitatively analyzed XMCD results. The obtained results are in very good quantitative agreement between the absorption-based (XMCD) and reflectivity-based (XRMR) techniques taking into account an ab initio calculated magnetooptic conversion factor for the XRMR analysis. Thus, it is shown that XRMR provides quantitative reliable spin depth profiles important for spintronic and spin caloritronic transport phenomena at this type of magnetic interfaces.

\end{abstract}

\maketitle

In the fields of spintronics \cite{Wolf2001} and spin caloritronics \cite{Bauer2012}, the generation and detection of pure spin currents play an essential role. Here, a common device is a non-magnetic material (NM) thin film used as spin current detector, which is grown on a ferromagnet (FM). The NM Pt is typically used for the conversion of the spin current into a transverse charge voltage via the inverse spin Hall effect \cite{Saitoh2006} because of its large spin Hall angle \cite{Hoffmann2013}. For a quantitative analysis however, one has to take other parasitic effects into account, which can occur due to the closeness of Pt to the ferromagnetic instability within the Stoner criterion description \cite{Stoner1938}. In layered systems of Pt in contact to a FM, the magnetic proximity effect (MPE) can generate a spin polarized interface within the Pt. This can lead to additional effects, e.g. a proximity-induced anomalous Nernst effect in spin Seebeck experiments \cite{Huang2012, Bougiatioti2017}  or a proximity-induced anisotropic magnetoresistance in spin Hall magnetoresistance studies \cite{Althammer2013}. It is therefore essential to investigate and understand the MPE in systems, which are used for the detection of pure spin currents. 

Furthermore, the influence of the MPE on spin-orbit torque (SOT) efficiencies is still under debate \cite{Peterson2018,Zhu2018,Moskaltsova2020}. Peterson \textit{et al.} \cite{Peterson2018} report an increase in the field-like SOT by nearly a factor of 4 at 20\,K, which is attributed to an increased magnetoresistance caused by the MPE at low temperatures. Contrarily, Zhu \textit{et al.} \cite{Zhu2018} claim negligible influence of the MPE on SOT efficiencies, whereas an enhancement of the MPE due to annealing was found.   

A commonly utilized effect for analyzing the MPE is x-ray magnetic circular dichroism\cite{Schutz1987} (XMCD), mostly studied in multilayers of e.g. Pt adjacent to FMs such as Fe\cite{Antel1999}, Co\cite{Schutz1990,Wilhelm2003a}, and Ni\cite{Wilhelm2000a}. Since the XMCD signal is an average over the whole Pt layer, usually a thickness variation is required to obtain quantitative values for the magnetic moment and the effective thickness of the spin polarized Pt interface layer \cite{Ruegg1991, Poulopoulos2001, Suzuki2005} using, e.g., fluorescence yield as detection mode, while there are also directly depth sensitive XMCD implementations on a single sample\cite{Amemiya2003}. Due to the MPE being an interface effect, a more natural approach for its detection is x-ray resonant magnetic reflectivity\cite{Macke2014b, Kuschel2015c} (XRMR). Here, changes in the magnetooptic absorption $\Delta \beta$ and magnetooptic dispersion $\Delta \delta$ make it directly sensitive to the magnetization density along the normal q-vector, hence the magnetic depth profile. 

However, additional spectroscopic measurements \cite{Seve1999} or calculations are necessary for a quantitative determination of the spin moments from an XRMR study as compared to XMCD with its sum rules \cite{Thole1992a, Carra1993, Chen1995}. Usually, the application of the sum rules to experimentally obtained XMCD spectra is necessary to quantify and translate the $\Delta \beta$ spectrum into the induced magnetic moment per atom. Another approach is the application of a magnetooptic conversion factor derived from ab initio calculations \cite{Kuschel2015c}, that has been used for obtaining magnetic moments per Pt atom at the interface from the magnetooptic depth profiles of the XRMR fitting. A systematic comparison of the Pt magnetic moments stemming from a detailed XMCD sum-rule analysis and from the magnetooptic depth profiles of XRMR together with the ab initio conversion factor within the same sample systems has not been presented so far. 

Therefore in this manuscript we tackle a detailed quantitative comparison between the spin magnetic moments measured by the absorption-based (XMCD) and reflectivity-based (XRMR) techniques for two sample stacks consisting of bilayers of Pt (3-4\,nm) adjacent to the FMs Fe and Co$_{33}$Fe$_{67}$ ($\sim$10\,nm). The samples were prepared by dc magnetron (co-)\-sputtering at room temperature from elemental targets onto MgO(001) substrates. The stoichiometry of the Co$_{33}$Fe$_{67}$ layer was verified by x-ray fluorescence spectroscopy. The first sample, Pt/Fe, is a standard combination when studying MPE \cite{Antel1999, Gepraegs2012, Kuschel2015c, Klewe2016, Kuschel2016}. The latter sample, Pt/Co$_{33}$Fe$_{67}$, was chosen due to its high magnetic moment and maximum MPE\cite{Bougiatioti2018}. The derived values for the maximum Pt magnetic moments in those studies\cite{Kuschel2016, Bougiatioti2018} were 0.5$\pm$0.1\,$\mu_{\textrm{B}}$ and 0.72$\pm$0.03\,$\mu_{\textrm{B}}$ per Pt atom for Pt/Fe and Pt/Co$_{33}$Fe$_{67}$, respectively, with a typical thickness of 1.1\,-\,1.2\,nm for the magnetic Pt layer.

\begin{figure}[H]
	\includegraphics[width=1\linewidth]{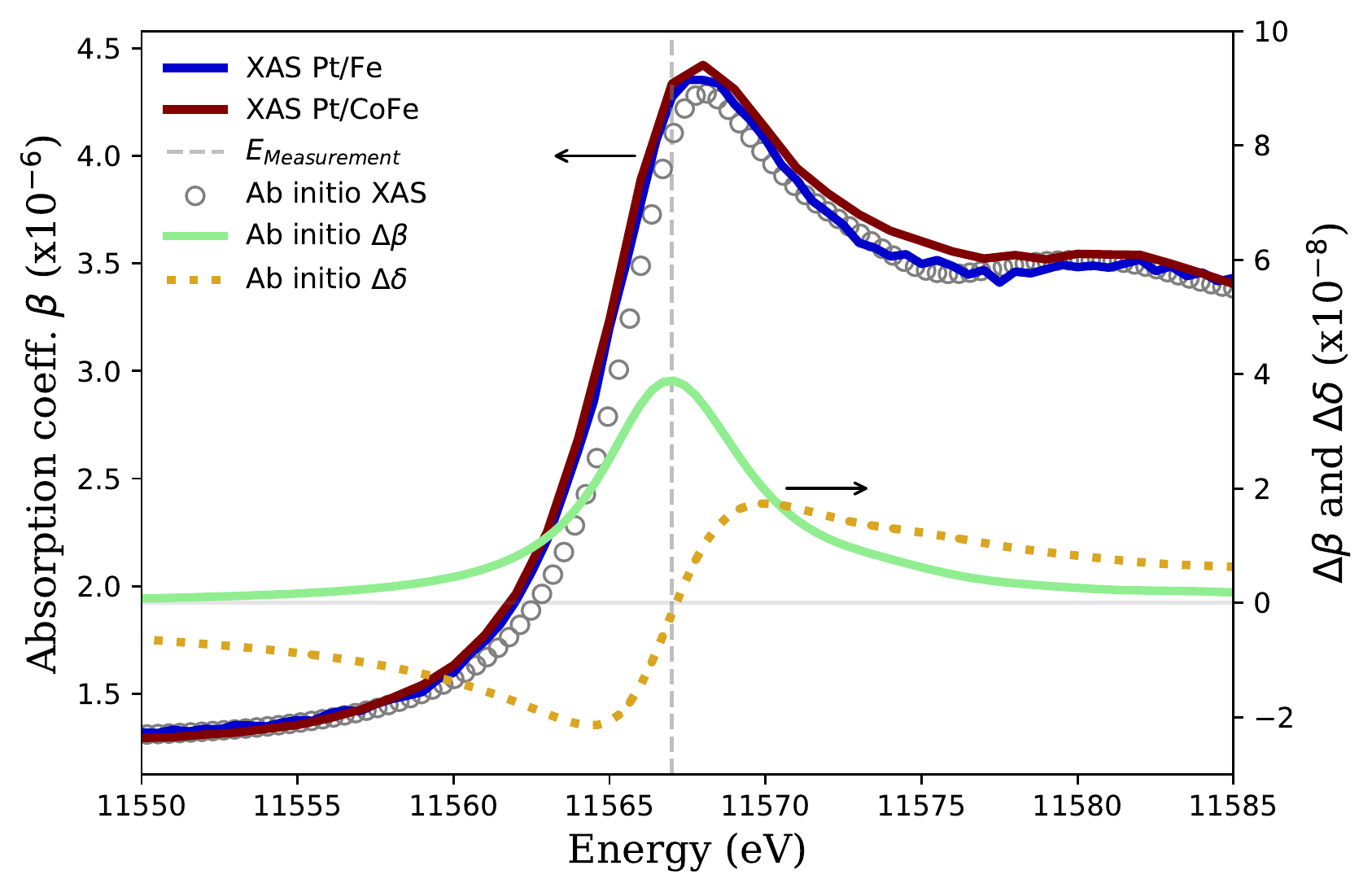}
	\caption{Calculated\cite{Kuschel2015c} and experimental Pt L$_{3}$ XAS spectra scaled to the absorption coefficient $\beta$ before and after the edge jump. Also shown are the calculated magnetooptic parameters $\Delta \delta$ and $\Delta \beta$\cite{Kuschel2015c}. The calculated spectra were shifted in energy to match the experimental XAS peak positions.}
	\label{XAS ab initio}
\end{figure}

The measurements were carried out at beamline P09 of the third-generation synchrotron at DESY. XRMR was measured in $\theta - 2\theta$ scattering geometry, with a fixed energy and fixed helicity of the circular polarized incident x-rays. The energy was chosen to be 1eV below \cite{Geissler2002,Kuschel2015c,Kuschel2016,Schutz1990} the peak of the Pt L$_{3}$ absorption edge\cite{Kuschel2015c}, called the whiteline, which was measured by x-ray absorption spectroscopy (XAS) as shown in Fig. \ref{XAS ab initio}. The corresponding XMCD measurements were also carried out with a fixed circular polarization of the x-rays. In order to magnetize the samples and generate a magnetic contrast for the measurements, the in-plane magnetic field was switched between parallel and antiparallel orientation with respect to the x-ray beam propagation, with a maximum applied magnetic field of $\pm$150\,mT. From the two spectra for each edge, Pt L$_{3}$ and L$_{2}$, the averaged absorption $\textrm{XAS} =\frac{I_{+}+I_{-}}{2}$ and the difference signal defined as $\textrm{XMCD} =I_{+}-I_{-}$, with the intensity $I_{\pm}$ for positive and negative magnetic field, respectively,  can be extracted. The measurements were repeated with the opposite x-ray helicity and the magnetooptic origin was verified by the sign change of the XMCD signal. The magnetic signals obtained for different photon helicities and magnetic field directions were combined to improve the signal-to-noise ratio and remove non-magnetic artefacts.

In Fig. \ref{XAS ab initio}, the measured XAS spectra for both samples together with the ab initio calculated absorption spectrum \cite{Kuschel2015c} are shown. They are relatively shifted in energy to match their peak position and scaled to the absorption coefficient $\beta$ before and after the edge step. Also shown are the ab initio calculated magnetooptic parameter $\Delta \beta$, together with $\Delta \delta$, which has been obtained by a Kramers-Kronig transformation of the $\Delta \beta$ spectrum. The incoming x-ray energy used for the XRMR measurements was particularly tuned to 1 eV below the white line peak maximum (grey vertical line in Fig. \ref{XAS ab initio}), to be in the maximum of the theoretical $\Delta \beta$ spectrum and without any $\Delta \delta$ contribution, thus, reducing the number of fit parameters. The XRMR measurements were also carried out twice, once with each circular polarization, and for each of them, the nonmagnetic x-ray reflectivity (XRR) $I=\frac{I_{+}+I_{-}}{2}$ and the asymmetry ratio $\Delta I = \frac{I_{+}-I_{-}}{I_{+}+I_{-}}$, are calculated. Finally, the averages for the two photon helicities and magnetic field directions are calculated to im-

\begin{figure}[H]
	\includegraphics[width=1\linewidth]{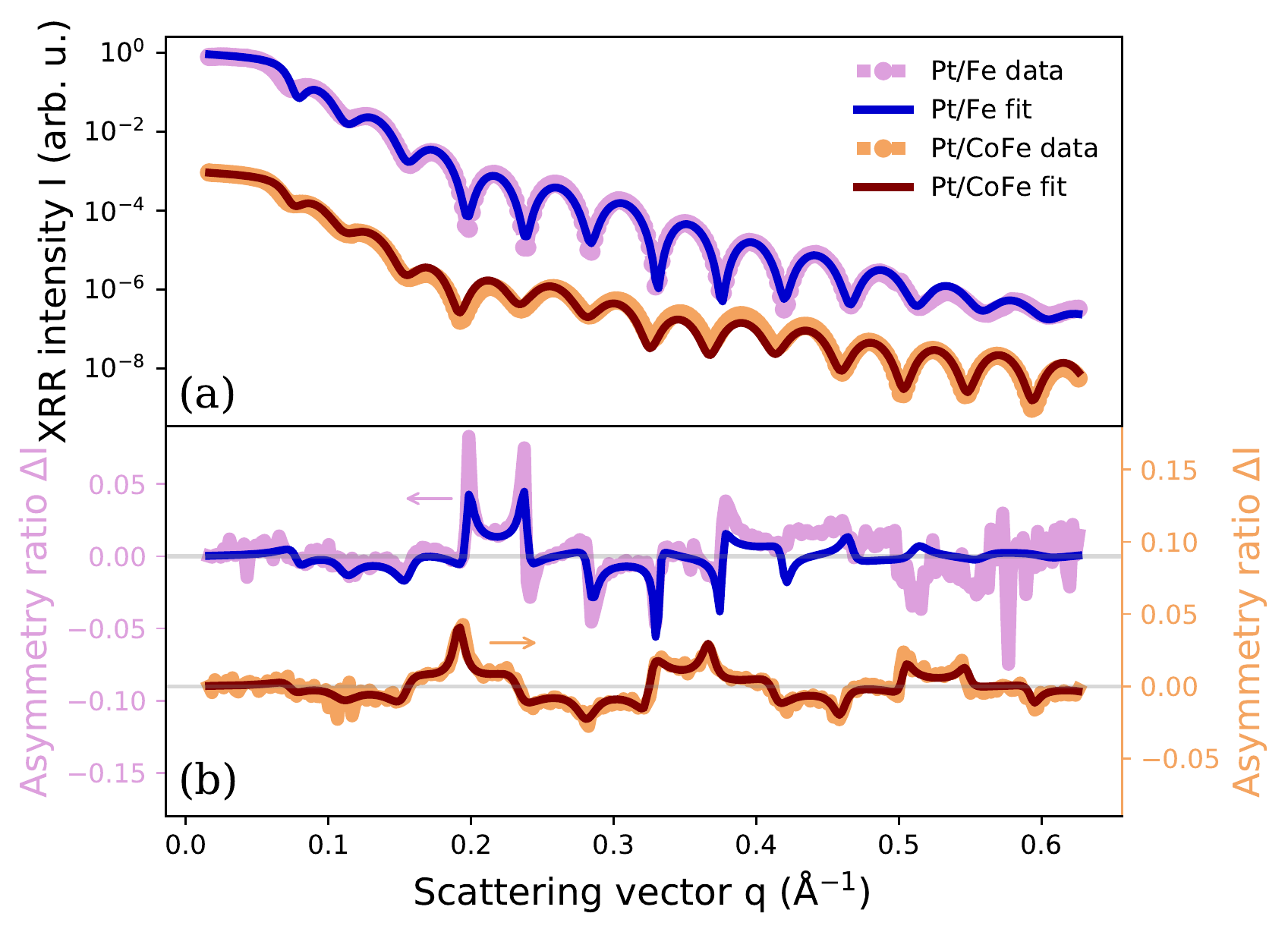}		
	\includegraphics[width=1\linewidth]{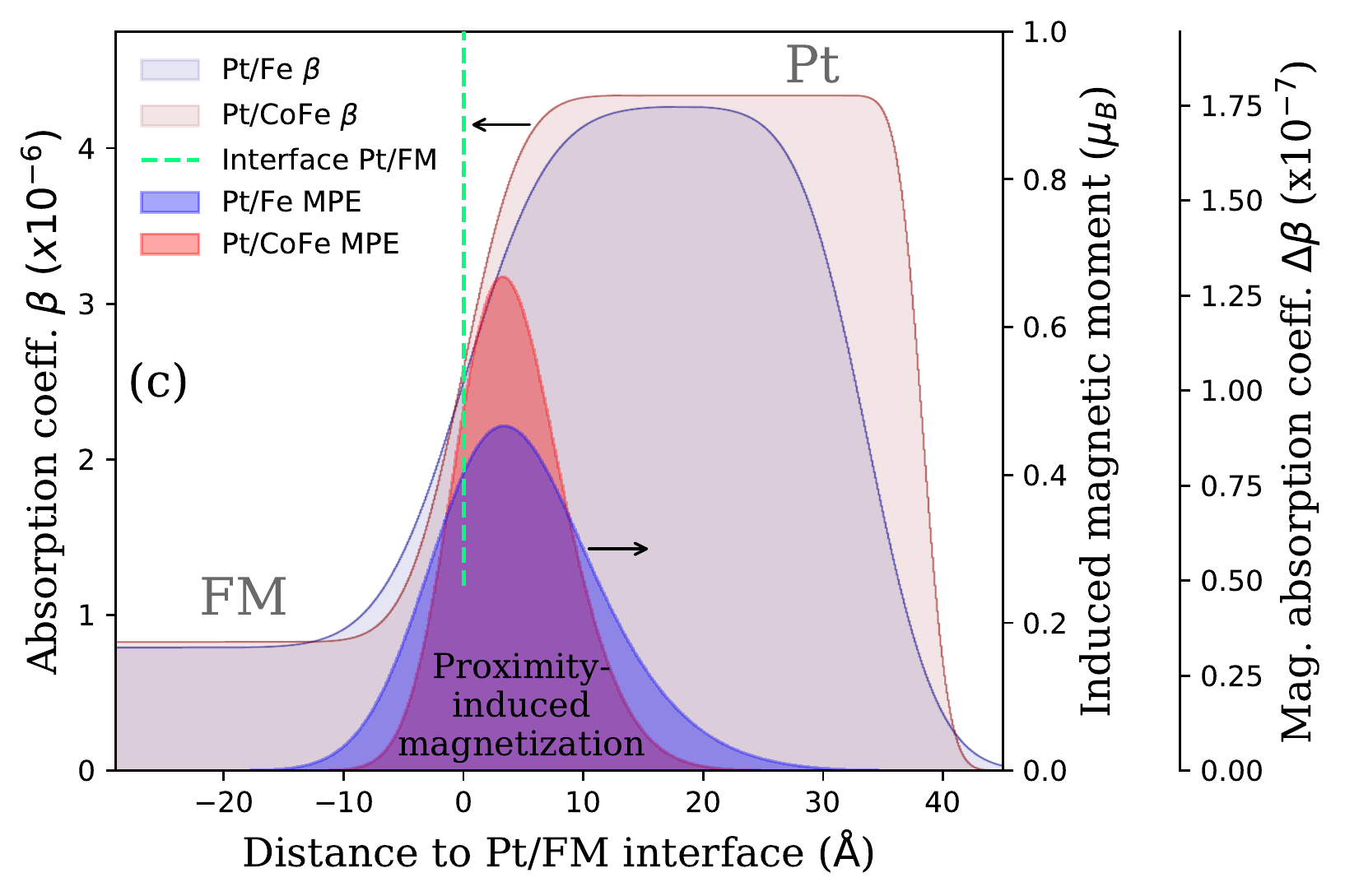}
	
	\caption{(a) Data and fits of the averaged resonant XRR curves. (b) Resulting asymmetry ratios calculated from the normalized difference of the two measurements with opposite x-ray helicity. (c) Modelled magnetooptic depth profiles of $\Delta \beta$ together with the derived  induced magnetic moment for the two samples (conversion factor taken as $\mu_{\textrm{spin}}^{\textrm{Pt}}=\Delta \beta \cdot 5.14\textrm{x}10^{6}\,\mu_{\textrm{B}}/\textrm{atom}$\cite{Kuschel2015c}).  The optical depth profiles obtained from the resonant XRR fit are also shown.}
	\label{XRMR}
\end{figure}

\noindent prove data quality. Further information on the two measurement techniques, the experimental details as well as the XRMR fitting procedure can be found in the Supplemental Material (including Ref.\cite{Strempfer2013, Macke2014a, Parratt1954, Nevot1980, Zak1990, Henke1993, Krieft2020}). 

Figures \ref{XRMR} (a) and (b) present the XRR curve and the XRMR asymmetry ratio, respectively, for both samples.
For fitting the data, the absorption $\beta$ of the Pt layer was taken from the XAS scans in Fig. \ref{XAS ab initio}, since this parameter showed to have the largest influence on the quantitative value of the resulting induced magnetic moments \cite{Klewe2016}. Having the structural and optical parameters fixed from fitting the XRR curves \cite{Klewe2016,Kuschel2015c}, the asymmetry ratio signal was modelled by the variation of a magnetooptic depth profile. Since the energy was chosen to minimize the influence of $\Delta \delta$ (see Supplemental Material for further information), only the spatial distribution and amplitude of $\Delta \beta$ was varied during fitting $\Delta I$. The modelled Gaussian depth profiles of $\Delta \beta$ were convoluted with the roughnesses of the Pt/FM interfaces, resulting in the Gaussian-like depth profiles shown in Fig. \ref{XRMR}(c), which correspond to the asymmetry ratio fits in Fig. \ref{XRMR}(b). The $\Delta \beta$ depth profile was converted into the induced Pt moment by the ab initio calculated conversion factor, see Fig. \ref{XRMR} (c), and for the maximum Pt moment the peak value of the Gaussian-like depth profile was taken.

For the Pt/Fe sample, we obtain a maximum spin magnetic moment of 0.47\,$\mu_{\textrm{B}}$ per Pt atom, which is in good agreement with the values previously reported for similar Pt/Fe samples \cite{Klewe2016, Kuschel2016}. Although a similar moment was found, the full width at half maximum (FWHM), an indication for the extension of the spin polarization, is slightly larger in the present work. We find a FWHM of 1.6\,nm within the 3.4\,nm Pt layer, whereas previously reported values are 1.1-1.2\,nm. This can be explained by a rather large roughness. When compared to the previously reported range of 0.45$\pm$0.10\,nm for the interfacial roughness for this kind of sample, the 0.55\,nm at the Pt/Fe interface of the present work is rather large. This probably results in a stronger intermixing of Pt and Fe and a therefore wider spin polarized Pt layer. 
For the Pt/Co$_{33}$Fe$_{67}$ sample, we obtain a maximum value for the spin magnetic moment of 0.67\,$\mu_{\textrm{B}}$ per Pt atom with a FWHM of 1\,nm within a 3.8\,nm Pt layer. Here, the roughness is smaller being 0.38\,nm, also indicating the interplay of interfacial roughness and spread of the proximity-induced magnetization. These quantitative XRMR results are summarized in Tab. \ref{Table_XMCD}. For both samples, only the spin contribution to the Pt magnetization is considered.

Since the asymmetry ratio data of the Pt/Co$_{33}$Fe$_{67}$ sample is better reproduced by the fit than for the Pt/Fe sample, especially after q = 0.4\,\AA{}$^{-1}$, alterations from the Gaussian model for the depth profile have been considered. However, the fit does not improve to a significant extent and the values at maximum as well as the FWHMs of the resulting $\Delta \beta$ profiles are within the given error range for that sample.   

\begin{figure}[]
	\includegraphics[width=1\linewidth]{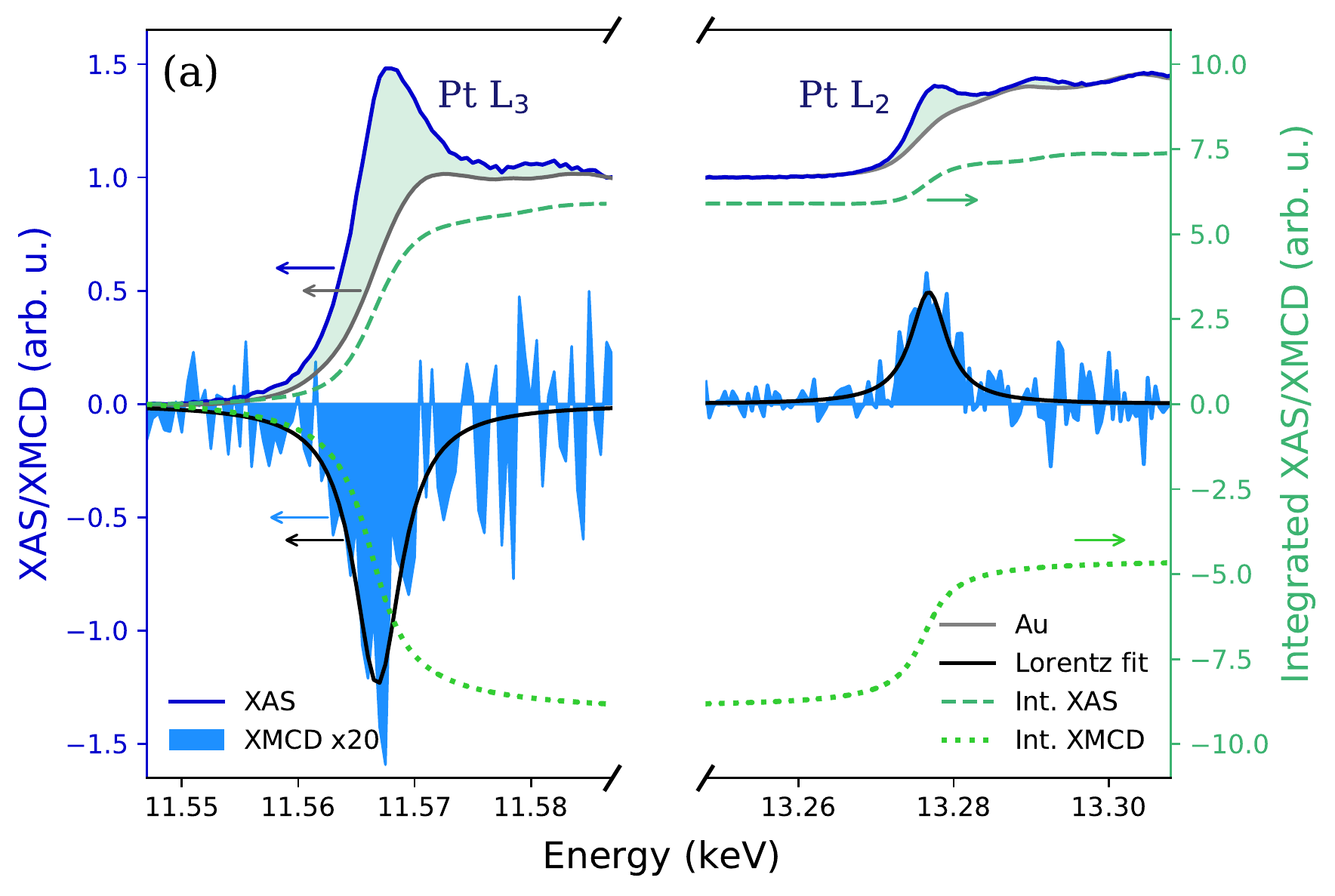}		
	\includegraphics[width=1\linewidth]{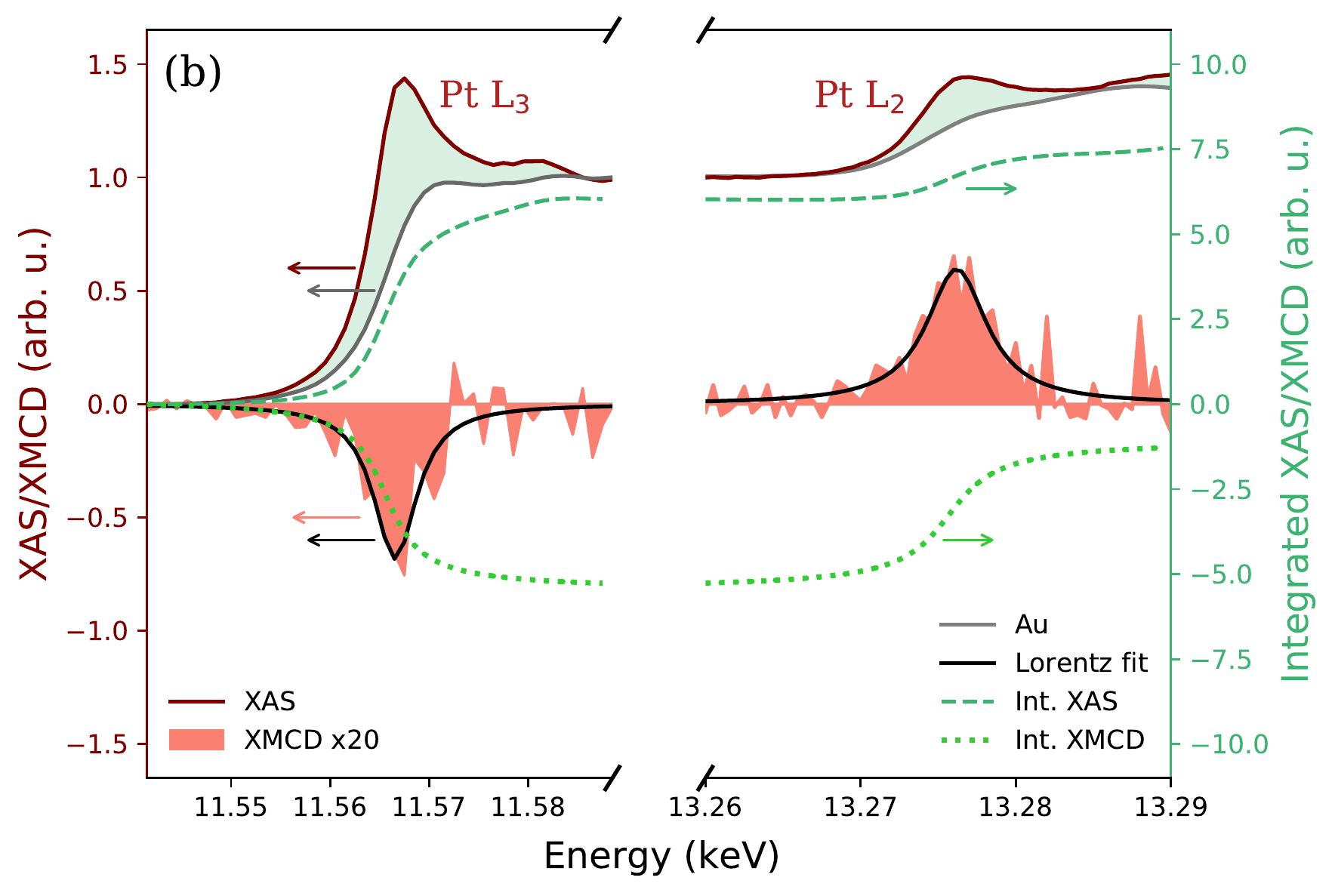}
	\caption{Experimental XAS for (a) the Pt/Fe sample and (b) the Pt/Co$_{33}$Fe$_{67}$ sample at the L absorption edges of Pt. The XAS data is averaged over the two x-ray helicities and magnetic field directions and the XMCD signal is scaled by a factor of 20 (filled area). Additionally shown is the Au reference absorption spectrum, shifted as described in the text. The integration of the difference between the Pt/FM XAS signal and the Au XAS as well as the integration of the XMCD signal is shown as dotted lines, for the XAS additionally indicated as the shaded area. The XMCD signal has been fitted with a Lorentzian to reduce the noise while integrating.}
	\label{XMCD}
\end{figure}

The XAS and XMCD spectra are presented in Fig. \ref{XMCD}. A clear dichroism can be seen for both samples at the Pt L$_{3}$ and L$_{2}$  edges. 
For the determination of the orbital and spin moments, the sum rules \cite{Thole1992a, Carra1993, Chen1995} were applied. As part of the data postprocessing, the pre-edge slope of the XAS spectra were corrected by a linear fit and the spectra were normalized to a L$_{3}$-L$_{2}$ edge jump ratio of $R=2.22$ as experimentally found for pure Pt\cite{Mattheiss1980}. For the removal of the continuum contribution, an experimentally obtained Au reference spectrum has been subtracted, following the procedure described in Ref.\cite{Grange1998}. Due to the generally small whiteline of Pt, the alternative approach of subtracting a step function, as typically done in the soft x-ray range, would lead to a large uncertainty contribution in the determination of the spin and orbital moments \cite{Wilhelm2001a}. 

XAS spectra of 0.2\,$\mu$m thick Au and Pt (Pt$_{\textrm{ref}}$) films, resembling the metallic state in terms of 5d holes, have been collected. The Au spectrum has been stretched and shifted in energy to match the extended x-ray absorption fine structure features of the Pt$_{\textrm{ref}}$ spectrum. By matching the near edge features of the Pt$_{\textrm{ref}}$ measurement to the Pt/FM samples, accounting for small variations in energy of the different measurements, the Au spectrum has been scaled as shown in Fig. \ref{XMCD}. The difference between the XAS of the Pt/FM and the Au reference, called $r_{\textrm{Pt/FM}}$, has been calculated as well as the difference between the XAS of Pt$_{\textrm{ref}}$  and of the Au reference ($r_{\textrm{Pt, ref}}$). The subtraction of the Au reference from the Pt$_{\textrm{ref}}$ with the relative difference in 5d holes\cite{Grange1998,Poulopoulos2001} $n_{\textrm{h}}^{\textrm{dif}}$ = 1.06 ($n_{\textrm{h}}^{\textrm{Pt, ref}} = 1.80$ and $n_{\textrm{h}}^{\textrm{Au}} = 0.74$) yields a scaling factor, which can be used to determine the number of holes for the Pt/FM samples\cite{Vogel1997a}: $n_{\textrm{h}}^{\textrm{Pt/FM}} = n_{\textrm{h}}^{\textrm{Au}} + r_{\textrm{Pt/FM}}$ $n_{\textrm{h}}^{\textrm{dif}} / r_{\textrm{Pt, ref}}$. 

With the integration of the L$_{3}$ XMCD signal $p$ and the total integrated XMCD signal, L$_{3}$ $+$ L$_{2}$, $q$, the sum rules can be written as
\begin{equation}
m_{\textrm{orb}}=\frac{2}{3}n_{\textrm{h}}^{\textrm{Pt/FM}}\frac{q}{r_{\textrm{Pt/FM}}}\quad;\quad  m_{\textrm{spin}}=n_{\textrm{h}}^{\textrm{Pt/FM}}\frac{\left(3p-2q\right)}{r_{\textrm{Pt/FM}}}.
\end{equation}  
The results can be found in Tab. \ref{Table_XMCD}. The procedure for the determination of the error bars for both methods can be found in the Supplemental Material (including Ref.\cite{Schoen2017}) together with a discussion on the orbital moments, which are neglected in the following analysis.

While the stoichiometry of the FM in the Pt/Co$_{33}$Fe$_{67}$ sample was chosen to have the largest spin moment as shown in a previous XRMR study \cite{Bougiatioti2018}, the obtained value from the XMCD analysis is smaller when compared to the Pt/Fe sample. This can be explained by looking at the results from the XRMR analysis displayed in Table \ref{Table_XMCD}. The Pt layer is slightly thicker for the Pt/Co$_{33}$Fe$_{67}$ sample and the FWHM of the spin polarized Pt is only two thirds of the value for the Pt/Fe sample. This coincides with the larger interfacial roughness of the latter sample. The obtained magnetic moments are relatively suppressed by the non-magnetic Pt, thus contributing less to the XMCD signal. Without a thickness variation of the Pt, the quantitative values derived from the XMCD sum-rule analysis give no information about the distribution of the magnetic atoms in the case of induced magnetism at the interface over a very small depth. 

The magnetic moment from the XMCD analysis gives an averaged moment over the complete Pt layer, while the XRMR depth profile considers a larger but at the Pt/FM interface located magnetization. For the quantitative comparison between the two x-ray techniques, we can compare the integrated area of the $\Delta \beta$ depth profile, scaled by the ab initio conversion factor, to the area of the total Pt depth profile, scaled to the XMCD spin moment. The differences are 4\% and 6\% for Pt/Fe and Pt/Co$_{33}$Fe$_{67}$, respectively, with the XRMR result being slightly larger for both samples. Furthermore, we can consider the spin magnetic moments obtained from the XRMR $\Delta \beta$ depth profile, when using the XMCD results instead of the ab initio calculation for the conversion to units of $\mu_{\textrm{B}}$. This would yield maximum spin magnetic moments of 0.45$\pm$0.14\,$\mu_{\textrm{B}}$ and 0.63$\pm$0.22\,$\mu_{\textrm{B}}$ for Pt/Fe and Pt/Co$_{33}$Fe$_{67}$, respectively. Despite the error bars being rather large, the obtained spin moment values are still in very good agreement, showing the quantitative reliability of the ab initio conversion factor. Further information on those calculations can be found in the Supplemental Material.

\begin{table}[]
	\begin{ruledtabular}
		\begin{tabular}{ccc}
			&Pt/Fe&Pt/Co$_{33}$Fe$_{67}$\\ \hline
			\rule{0pt}{3.3ex}
			$t_{\textrm{\textrm{Pt}}}$&3.37$\pm$0.05\,nm& 3.84$\pm$0.05\,nm\\
			\rule{0pt}{3.3ex}
			$t_{\textrm{MPE}}$&1.56$\pm$0.10\,nm& 1.04$\pm$0.10\,nm\\
			\rule{0pt}{3.3ex}
			$\sigma_{\textrm{Pt/FM}}$&0.55$\pm$0.05\,nm& 0.38$\pm$0.05\,nm\\
			\rule{0pt}{3.3ex}
		    $m_{\textrm{orb}}^{\textrm{XMCD, lim}}$&$<$0.05\,$\mu_{\textrm{B}}$&$<$0.015\,$\mu_{\textrm{B}}$\\
		    \rule{0pt}{3.3ex}
			$m_{\textrm{spin}}^{\textrm{XMCD}}$&0.22$\pm$0.05\,$\mu_{\textrm{B}}$&0.18$\pm$0.04\,$\mu_{\textrm{B}}$\\
			\rule{0pt}{3.3ex}
			$m_{\textrm{spin}}^{\textrm{XRMR}}$&0.47$\pm$0.10\,$\mu_{\textrm{B}}$&0.67$\pm$0.10\,$\mu_{\textrm{B}}$\\
			\rule{0pt}{3.3ex}
			$m_{\textrm{spin}}^{\textrm{XMCD, scaled}}$&0.45$\pm$0.14\,$\mu_{\textrm{B}}$&0.63$\pm$0.21\,$\mu_{\textrm{B}}$\\	
		\end{tabular}
	\end{ruledtabular}
	\caption{\label{Table_XMCD}Results from XRMR (including $\Delta \beta$-to-magnetic-moment conversion) and XMCD sum-rule analysis. Displayed are the thickness of the Pt layer $t_{\textrm{Pt}}$, the FWHM of the magnetic part of Pt $t_{\textrm{MPE}}$ and the interfacial roughness $\sigma_{\textrm{Pt/FM}}$ as obtained from XRMR. The XMCD sum-rule analysis yields the spin moment $m_{\textrm{spin}}^{\textrm{XMCD}}$ and an upper limit for the orbital moment  $m_{\textrm{orb}}^{\textrm{XMCD, lim}}$. The value $m_{\textrm{spin}}^{\textrm{XRMR}}$ is derived from the XRMR analysis using the ab initio conversion factor, while $m_{\textrm{spin}}^{\textrm{XMCD, scaled}}$ is obtained by using the XMCD result for the scaling of the $\Delta \beta$ depth profile.}
\end{table}

Being the first comparison of this kind, we can only call for further such multi-modal studies of interfacial magnetism, both for systematic quantitative checks of consistency between XRMR and XMCD as well as for the inherent complementarity of these two probes.

For samples with very high interfacial roughness or non conducting FMs, where no MPE is expected, and its presence is caused by interdiffusion due to the sample preparation, the differences in the quantitative results seem to get larger as found for the inverted Y$_{3}$Fe$_{5}$O$_{12}$/Pt structure\cite{Gepraegs2020}. This could be caused by the change of the atomic surrounding of the Pt, limiting the applicability of the calculated ab inito conversion factor or sum-rule analysis, or other factors which cannot be described by at least one of the two techniques without further investigations. 

Concluding, it was shown within this work, that the quantitative results of the XRMR analysis, including the $\Delta \beta$-magnetic-moment conversion factor, are very much comparable to the ones obtained by XMCD and a sum-rule analysis for the given kind of samples. While XRMR and XMCD are complementary methods and a multi-modal approach is surely favourable in the case of unknown sample types, it has been shown for the given sample type, that XRMR can be used as a stand-alone quantitative method. Provided that an ab initio factor can be calculated and has been confirmed, XRMR can be used in the analysis of magnetic depth profiles, as shown here for Pt and samples involving low interfacial roughness, without the need of conducting additional XMCD measurements to obtain reliable results. The increase in magnetization in the NM due to the MPE must be known for a quantitative analysis of phenomena such as SOT efficiencies and spin Hall angle measurements. Especially, when comparing samples with slight variations in the thickness of the NM and/or the spin polarized layer (e.g. due to different growth conditions) or samples with more than one interface of interest (e.g. trilayers with asymmetric MPE \cite{Inyang2019, Mukhopadhyay2020}), any XMCD result alone for instance can only reveal changes in the overall averaged magnetic moment value and not the differences in the magnetic moment distribution. Exactly the former case has been highlighted in our presented study.
However, both information, the spatial distribution of magnetic moments as well as its quantification, are now available and confirmed using XRMR in combination with the reliable $\Delta \beta$-magnetic-moment conversion factor for Pt being $\mu_{\textrm{spin}}^{\textrm{Pt}}=\Delta \beta\cdot5.14\textrm{x}10^{6}$$\,\mu_{\textrm{B}}/\textrm{atom}$.

\section*{Supplementary Material}

See Supplemental Material for additional information about the measurement techniques, XRMR fitting procedure, calculations for the quantitative comparison of the XRMR and XMCD results, the determination of the error bars, Kramers-Kronig transformation of the XMCD signal, and a discussion on the orbital moments as obtained from the XMCD sum-rule analysis.\\

We acknowledge DESY (Hamburg, Germany), a member of the Helmholtz Association HGF, for the provision of experimental facilities. Parts of this research were carried out at beamline P09 at PETRA III. We would like to thank G\"unter Reiss for making available the laboratory equipment in Bielefeld and acknowledge financial support by the Deutsche Forschungsgemeinschaft (DFG) within the grant RE 1052/42-1.\\ 

The data that support the findings of this study are available from the corresponding author upon reasonable request.

\nocite{*}
\bibliography{Comparison_XRMR_XMCD}

\end{document}